\documentclass[aps,twocolumn,pra]{revtex4-1}
\usepackage{newlfont}
\usepackage{amssymb}
\usepackage{amsfonts}
\usepackage{amsmath}
\usepackage{wasysym}
\usepackage{graphicx}
\usepackage{bm}

\usepackage{epsfig}

\usepackage{amsthm}

\begin{document}

\title{Shared Purity of Multipartite Quantum States}

\author{Anindya Biswas, Aditi Sen(De), and Ujjwal Sen}

\affiliation{Harish-Chandra Research Institute, Chhatnag Road, Jhunsi, Allahabad 211 019, India}


\begin{abstract}

Fidelity plays an important role in measuring distances between pairs of quantum states, of single as well as 
multiparty systems. Based on the concept of fidelity, we introduce a physical quantity, shared 
purity, for arbitrary pure or mixed quantum states of shared systems of an arbitrary number of parties in arbitrary dimensions. 
We find that it is different from quantum correlations. However, we prove that 
a maximal shared purity between two parties excludes any shared purity of these parties with a third party, thus ensuring its quantum nature.
Moreover, we show that all generalized GHZ states are monogamous, while all generalized W states are non-monogamous with respect to this measure.
  We apply the quantity to investigate the quantum XY spin models, and observe that it can faithfully 
detect the quantum phase transition present in these models. We perform a finite-size scaling analysis and find the 
scaling exponent for this quantity. 

\end{abstract}

\maketitle

\section{Introduction}

In the last two decades, various quantum information protocols like quantum teleportation~\cite{Chbennett1}, super dense coding~\cite{Chbennett2},
and entanglement-based quantum cryptography~\cite{Akekert} were discovered in which quantum mechanics is used to achieve higher efficiencies than their classical 
counterparts. Most of these protocols use entanglement~\cite{HHHH-RMP} as the resource. However, notable exceptions exist. 
In particular, there exists  orthogonal product states which are not perfectly distinguishable by measurements based on
local operations and classical communication~\cite{Chbennett3} (cf.~\cite{Aperes}).
In deterministic quantum computation
with one qubit~\cite{Eknill}, one achieves nonclassical efficiencies by employing shared quantum states with no entanglement. 
A protocol for secure deterministic communication without entanglement was also proposed and experimentally realized~\cite{Mlucamarini,Acere}. 

In a bid to understand such phenomena, measures of quantum correlation that are independent of the entanglement-separability
paradigm have been proposed~\cite{Kmodi}. Such measures, generally referred to as information-theoretic quantum correlation measures,
have been able to provide important successes in understanding these
phenomena. However, there is still a lot of ground that remains to be covered, and moreover, there are intriguing controversies
that have been generated~\cite{Kmodi}. This has led us to believe that it is important to identify 
concepts of
quantum shared states that are independent of quantum correlations.

In prospecting for possible physical quantities that lead to nonclassical phenomena by using shared states of quantum systems, 
we note that the maximal advantage in most quantum information protocols is achieved for \emph{pure} shared states. 
Pure quantum states, which in general play an important role
in quantum mechanics, have vanishing entropy, implying that the full information of the system is available, so that 
there always exists a measurement strategy for which the system leads to an outcome with unit probability. 


For a quantum state of a single system, the fidelity of that state to a pure state, after maximization over all pure states, is the 
purity of that state. It gives the maximal Born probability that can be obtained in any outcome of any unit-rank quantum measurement
on the given state. In case this quantum system (pure or mixed) consists of two or more subsystems, and the maximization is still performed
over all pure states of the entire system, the maximal fidelity can be referred to as the ``global fidelity''. 
Similarly, we
introduce a ``local fidelity'', that  measures the maximal Born probability which can be obtained in any product state outcome 
of any unit-rank measurement on the given state. Specifically, the maximization has now to be performed over a suitably chosen
set of pure product states. Depending on the set of product states in the  maximization, there is a hierarchy 
of such local fidelities. ``Shared purity'' is the difference between the global and a local fidelity. Despite its similarity in 
form with information-theoretic quantum correlation measures like quantum discord~\cite{Hollivier} and quantum work deficit~\cite{Joppenheim}, in that 
the latter are also differences between two state functions, we will show that shared purity  quantifies a property of shared quantum systems that is 
different from entanglement as well as information-theoretic measures. 
After introducing the measure, we discuss its properties and in particular show that
 for pure states, it reduces to geometric measures of entanglement \cite{Ashimony,Asde}. However, for mixed states, it is different from any quantum 
correlation measure. We provide analytical closed forms of shared purity for several classes of multiparty mixed quantum states. 
We then address the question of monogamy for this measure and prove that the quantity is qualitatively monogamous in that a maximal shared purity between 
two parties excludes any shared purity of these parties with a third one. It is to be noted that classical correlations are not monogamous, 
even qualitatively, and hence the monogamous nature of shared purity ensures the quantumness of the property. 
We further show that according to the shared purity, W-class~\cite{ZHG, Wdur} 
states are (relatively) more non-monogamous than the Greenberger-Horne-Zeilinger-class (GHZ-class) states~\cite{Wdur,GHZ}, among three qubit pure states.

Moreover, we apply the measure to quantify properties of the quantum anisotropic XY spin model. It has been shown that bipartite 
as well as multipartite entanglement measures can detect the quantum phase transition present in this model~\cite{Mlewenstein,Lamico}. 
We show that shared purity is also a faithful detector for identifying the phase transition. We obtain the finite-size scaling in
 this model for shared purity, and find it to be different from that for the entanglement measure called 
concurrence~\cite{Chbennett4,Aosterloh}, and for the information-theoretic quantum correlation measure called quantum discord \cite{Hollivier}.

The paper is organized as follows. In Sec.~\ref{sp}, we define shared purity and discuss some of its properties. 
In Sec.~\ref{ms}, we evaluate the shared purity for 
various classes of mixed states. In Sec.~\ref{mg}, we address the question of monogamy for shared purity. We apply the
measure to the anisotropic XY model and detect the quantum phase transition in this system in Sec.~\ref{qpt}. 
Finally, we present a discussion in Sec.~\ref{concl}.

\section{Shared Purity: definition \& properties}
\label{sp}
The fidelity is a very useful concept in quantum information, or more generally, in quantum mechanics. It quantifies the closeness of two quantum 
states, when at least of them is pure. It can be directly connected to the geometric distances in quantum mechanics like the 
Fubini-Study metric~\cite{Janandan} or the Bures metric~\cite{bures}. We define the ``global fidelity'' of an $N$-party arbitrary
(pure or mixed) quantum state, $\rho_{1\ldots N}$, on $\cal{H}=\mathbb{C}^{d_1}\otimes\ldots\otimes \mathbb{C}^{d_N}$, as 
\begin{equation} 
 F_G=\displaystyle \max_{\lbrace|\phi\rangle_{1\ldots N}\in \mathcal{H}\rbrace} {_{1\ldots N}\langle\phi|\rho_{1\ldots N}|\phi\rangle_{1\ldots N}},
\end{equation}
where the maximization is performed over all elements (pure states) of $\mathcal{H}$. The global fidelity quantifies the 
minimum distance of an arbitrary multipartite (pure or mixed) state from  pure quantum states, {\it i.e.} from states with vanishing global entropy. 
It measures the lack of disorder present in the system. In case of pure states, the minimum is attained by itself
and therefore it is unity for all pure states in arbitrary dimensions. It is also possible to 
obtain $F_G$ in closed form for arbitrary mixed states. In particular, we have the following theorem for arbitrary mixed states.\\\\
\noindent\textbf{Theorem 1.} \emph{For an arbitrary mixed state $\rho_{1\ldots N}$, $F_G$ is the largest eigenvalue in the spectrum of the state}.\\
\noindent \texttt{Proof:} Let us write $\rho_{1\ldots N}$ in spectral decomposition as 
 $\rho_{1\ldots N}=\displaystyle \sum_i \lambda_i|e_i\rangle\langle e_i|$, 
where $\lbrace|e_i\rangle\rbrace$ forms an orthonormal basis spanning $\mathcal{H}$. On the other hand, given a basis $\lbrace|e_i\rangle\rbrace$, 
any pure state can be written as $|\phi\rangle_{1\ldots N}=\displaystyle \sum_i a_i|e_i\rangle$,
where $a_i$ are complex numbers, with $\displaystyle \sum_i |a_i|^2=1$.
Therefore, $F_G=\displaystyle \max_{a_i}\sum_i |a_i|^2\lambda_i$. Let $\lambda_r$ be the largest eigenvalue in the spectral decomposition.
Assuming $a_r=1$ and $a_{i,i\ne r}=0$, we get $F_G\geq \lambda_r$. But, $F_G\leq\displaystyle\max_{a_i}\sum_i|a_i|^2\lambda_r=\lambda_r$,
since $\lambda_i\leq\lambda_r~~\forall i$ and $\displaystyle \sum_{i}|a_i|^2=1$. Therefore $F_G=\lambda_r$, and we have the theorem. \hfill $\blacksquare$\\

A class of pure states of multiparty systems that are drastically different, in terms of their quantum information hierarchy,
from other pure states is the class of pure product states. It is therefore interesting to find out how much of the global 
fidelity can already be attained by using such ``local'' states. Correspondingly, we define the ``local fidelity'', of the $N$-party 
quantum state $\rho_{1\ldots N}$ as
\begin{equation} 
 F_L=\displaystyle \max_{\lbrace|\phi\rangle_{1\ldots N} \in S\rbrace} 
{_{1\ldots N}\langle\phi|\rho_{{1}\ldots{N}}|\phi\rangle_{1\ldots N}},
\end{equation}
where the maximization is carried out over a certain set $S$, of pure product states. Depending on the set of product states, over which 
the optimization is carried out, there can be a hierarchy of local fidelities. There are two extreme cases that we mention explicitly.
 One is when the maximization is carried out over the set consisting of the fully separable states, {\it i.e.}, states of the form 
$|\phi^{(1)}\rangle_1 \otimes \ldots \otimes |\phi^{(N)}\rangle_N$. This is the class for which the entropies of each party vanish. 
We call this set as $S_L$, and for simplicity, denote the local fidelity by \(F_L\) when the optimization is performed over \(S_L\). 
An important class of multiparty states is the one consisting of genuinely multiparty entangled states, which are 
states that are entangled across every bipartition. The local fidelity optimized over the set, \(S_{n-gen}\), of states which are not genuinely multiparty entangled
will be denoted by $F_L^{n-gen}$. We now show that $F_L^{n-gen}$ can be analytically related to Schmidt coefficients in case of pure states.\\\\
\noindent\textbf{Theorem 2.} \emph{For an arbitrary pure $N$-party state $|\psi\rangle_{1\ldots N}$, $F_L^{n-gen}$ is the square of the maximal 
Schmidt coefficient among all bipartitions}.\\
\noindent \texttt{Proof:} For the pure state $|\psi\rangle_{1\ldots N}$, $F_L^{n-gen}$ reduces to 
$F_L=\displaystyle \max_{\lbrace|\phi\rangle_{1\ldots N}\in S_{n-gen}\rbrace} |\langle\phi|\psi\rangle|^2=1-\cal{E}$, where $\cal{E}$ is the generalized 
geometric measure~\cite{Asde} (c.f.~\cite{Ashimony}). As was shown in Refs.~\cite{Asde,Abiswas}, 
${\cal E}(|\psi\rangle) = 1 - \max \{\lambda^2_{{\cal A}: {\cal B}} | {\cal A} \cup {\cal B} = \{1,\ldots,N\}, {\cal A} \cap  {\cal B} = \emptyset\}$, 
where $\lambda_{\cal{A}:\cal{B}}$ is the maximal Schmidt coefficient in the $\cal{A}:\cal{B}$ bipartition. Hence, the theorem. \hfill $\blacksquare$\\

The definitions of the global and local fidelities imply that the local fidelity can never surpass in value  the global one. There is therefore
a part of the global fidelity that \emph{may not} be accounted for by the shared state's lack of disorder as seen by its local parties. This portion of 
the global fidelity is therefore present due to the fact that the state is shared, and can be quantified by the difference between the global
and local fidelities. We call the quantity as the ``shared purity''. Depending on the set of product states over which the optimization is carried 
out in defining the local fidelity, there can be a hierarchy of shared purities. For example, if the set consists of states which are not 
genuinely multiparty entangled, then the corresponding shared purity can be called the ``genuine shared purity''. We denote it by $S_P^{n-gen}$:
\begin{equation}
 S_P^{n-gen}=F_G-F_L^{n-gen}.
\end{equation}
On the other hand, if the optimization in the local fidelity is carried out over the set,  $S_L$, of fully separable states, we denote the corresponding shared purity
by $S_P$:
\begin{equation}
 S_P=F_G-F_L.
\end{equation}
Let us now discuss some properties of shared purity.\\\\
%
%
%
%
%
\noindent{\bf Property 1.} The shared purity vanishes for pure product states of the form \(|\psi_1\rangle \otimes \ldots \otimes |\psi_N\rangle\).\\\\
\noindent\textbf{Property 2.} For an arbitrary $N$-party pure state $|\psi\rangle_{1\ldots N}$ in arbitrary dimensions, the shared purity is a geometric measure of entanglement.\\
\noindent \texttt{Proof:} For arbitrary pure states, $F_G=1$. Now, the local purity is given by 
$\displaystyle \max_{\lbrace|\phi\rangle_{1\ldots N} \in S\rbrace}|_{1\ldots N}\langle\phi|\psi\rangle_{1\ldots N}|^2$, where 
\(|\phi\rangle_{1\ldots N}\) belongs to a suitably chosen set, \(S\), of pure product states. It is known that 
the quantities \(1 - \displaystyle \max_{\lbrace|\phi\rangle_{1\ldots N} \in S\rbrace}|_{1\ldots N}\langle\phi|\psi\rangle_{1\ldots N}|^2\)
are entanglement monotones \cite{Ashimony, Asde, Abiswas}. In particular, if \(S =S_L\), the quantity is called the geometric 
measure of entanglement \cite{Ashimony}, and if \(S= S_{n-gen}\), it is called the 
generalized geometric measure \cite{Asde, Abiswas}. \hfill \(\blacksquare\).\\  
As we will see in the next section, in the case of mixed states, the shared purity cannot be identified with any entanglement measure, or indeed any quantum correlation.\\\\
\noindent{\bf Property 3.} The shared purity is invariant under local unitary operations.\\\\
\noindent{\bf Theorem 3:} For an arbitrary bipartite (pure or mixed) state, on $\mathbb{C}^{d_1} \otimes  \mathbb{C}^{d_2}$, the minimum  value attained 
by $F_L$ is 
\(\lambda_r/d\), where \(d=\min\{d_1,d_2\}\), and \(\lambda_r\) is the largest eigenvalue in the spectrum of \(\rho\).\\
\noindent \texttt{Proof:} 
We have 
\begin{eqnarray}
F_L(\rho) = \max_{\lbrace|\phi\rangle \in S_L\rbrace} \langle \phi | \rho |\phi\rangle \nonumber \\
          = \max _{\lbrace|\phi\rangle \in S_L\rbrace} \sum_i \lambda_i |\langle \phi | e_i\rangle|^2,  
\end{eqnarray}
where \(\sum_i p_i |e_i\rangle \langle e_i|\) is a spectral decomposition of the bipartite quantum state \(\rho\).
Therefore, 
\begin{equation}
 F_L(\rho) \geq \max _{\lbrace|\phi\rangle \in S_L\rbrace}  \lambda_r  |\langle \phi | e_r\rangle|^2.
\end{equation}
The property follows from the fact that \(F_L \geq \frac{1}{d}\) for any pure state in \(\mathbb{C}^{d_1} \otimes  \mathbb{C}^{d_2}\). \hfill $\blacksquare$\\\\
\noindent \textbf{Corollary 3.1.} For an arbitrary bipartite (pure or mixed) state, on $\mathbb{C}^{d_1} \otimes  \mathbb{C}^{d_2}$, the maximum  value attained 
by \(S_P\) is 
\(\lambda_r(1-1/d)\), where \(d=\min\{d_1,d_2\}\), and \(\lambda_r\) is the largest eigenvalue in spectrum of \(\rho\).\\
\noindent \texttt{Proof:} This follows from Theorems 1 and 3 and the fact that 
\(\sup (f+g) \leq \sup f + \sup g\), for two bounded real-valued functions \(f\) and \(g\), defined on the same domain of definition. \hfill $\blacksquare$\\\\
Since \(\lambda_r \leq 1\), we have that  for an arbitrary bipartite (pure or mixed) state, on $\mathbb{C}^{d_1} \otimes  \mathbb{C}^{d_2}$, the maximum  value attained 
by \(S_P\) is 
\(1-1/d\), where \(d=\min\{d_1,d_2\}\). Moreover, this maximal value is attained by a maximally entangled state. 

\section{Mixed states}
\label{ms}

In this section, we will consider the shared purity for nonpure states, and in particular reach the conclusion that shared purity is different
 from any quantum correlation measures. 
It will be shown to be different from any measure of quantum correlation of the entanglement-separability paradigm~\cite{HHHH-RMP}, like entanglement 
of formation~\cite{Chbennett4}, relative entropy of entanglement~\cite{Vvedral}, logarithmic negativity~\cite{Gvidal}, 
etc., and also from 
any such measure of the information-theoretic paradigm~\cite{Kmodi}, like quantum discord~\cite{Hollivier}, quantum work-deficit~\cite{Joppenheim}, etc.

We begin by proving the following result for 
classically correlated states.\\\\
\noindent\textbf{Property 4.} For classically correlated states, the global and local fidelities are equal.\\
\noindent \texttt{Proof:} In general, the classically correlated state is given by
\begin{equation}
\rho_{1\ldots N}=\displaystyle \sum_{i_1,\ldots,i_N} p_{i_1\ldots i_N}|i_{1}\rangle \langle i_1|\otimes\ldots\otimes |i_{N}\rangle \langle i_N| \nonumber
 \end{equation}
where $\lbrace |i_j\rangle \rbrace_{i_j=1}^{d_j}$, for \(j=1,\ldots, N\),  forms a mutually orthonormal set of vectors. 
By using Theorem 1,
we get that $F_G$ is the largest eigenvalue, say $p_{i_1\ldots i_N}$, corresponding to the spectral vector,  $|i_1\rangle\otimes\ldots\otimes|i_N\rangle$.
Hence the maximum in the  global fidelity is attained in a completely product state. Hence, all local fidelities can also be attained in the same state, yielding the same value as the global one. 
\hfill $\blacksquare$\\\\
\noindent\textbf{Property 5.} For a state of the form \(\rho_1 \otimes \ldots \otimes \rho_N\) on \(\mathbb{C}^{d_1}\otimes \ldots \otimes \mathbb{C}^{d_N}\),  shared purity vanishes.\\\\

Below, we will find that shared purity can be nonzero for separable states. However, it can also be vanishing for entangled states. These twin facts imply that shared purity is conceptually 
different from quantum correlation measures, even the information-theoretic ones. 

\subsection{Bipartite and multipartite mixed states: some examples}
%
%
%

We will now investigate the behavior of shared purity for paradigmatic classes of mixed states. 
 \subsubsection{Admixtures of a Bell state with a pure product state}
 \label{gaaner-onek-sur-samudrer-jole}
Consider the state 
\begin{equation}
\rho_{ent}=p|00\rangle\langle00|+(1-p)|\psi^-\rangle\langle\psi^-|, \nonumber
\end{equation}
where $|\psi^-\rangle=\frac{1}{\sqrt{2}}(|01\rangle-|10\rangle)$ and \(0\leq p\leq 1\). \(|0\rangle\) and \(|1\rangle\)
form the computational basis of a single qubit.  
 The state is entangled for any value of $p<1$.
Since the state is already written in its spectral decomposition, $F_G$ is equal to the coefficient of the largest spectral
 component, {\it i.e.} $F_G=\max \{p,1-p\}$.
 
Now, we present some of the steps involved in the calculation of $F_L$. The states $|\phi_A\rangle$ and $|\phi_B\rangle$, of the product states \(|\phi_A\rangle \otimes |\phi_B\rangle\)
over which $F_L$ is maximized, are given by
\begin{eqnarray}
|\phi_A\rangle &=& e^{i\phi_1}\cos\frac{\theta_1}{2}|0\rangle+e^{i\phi^{'}_1}\sin\frac{\theta_1}{2}|1\rangle, \nonumber \\
|\phi_B\rangle &=& e^{i\phi_2}\cos\frac{\theta_2}{2}|0\rangle+e^{i\phi^{'}_2}\sin\frac{\theta_2}{2}|1\rangle. \nonumber
\end{eqnarray}
Therefore,
\begin{equation}
 F_L=\displaystyle \max_{\theta_1,\theta_2} \left\lbrace p\cos^2\frac{\theta_1}{2}\cos^2\frac{\theta_2}{2}+\frac{1-p}{2}\sin^2\left(\frac{\theta_1+\theta_2}{2}\right)\right\rbrace, \nonumber
\end{equation}
after we have already maximized $F_L$ over the set of variables $\lbrace\phi_1,\phi^{'}_1,\phi_2,\phi^{'}_2\rbrace$
by setting $\cos(\phi^{'}_1+\phi_2-\phi_1-\phi^{'}_2)=-1$. Optimizing $F_L$ with respect to $\lbrace\theta_1,\theta_2\rbrace$ leads us to the 
following three conditions -- $\cos\frac{\theta_1}{2}=0$ or $\cos\frac{\theta_2}{2}=0$ or $\sin\frac{\theta_1-\theta_2}{2}=0$. Each of the first
two conditions extremizes $F_L$ to $\frac{1-p}{2}$ while the third gives 
\begin{eqnarray}
 F_L &=& \frac{(1-p)^2}{2-3p},~~~~~~p\leq\frac{1}{2}, \nonumber \\
     &=& p,~~~~~~~~~~~~~~~ p\ge \frac{1}{2} \nonumber.
\end{eqnarray}
Therefore the third condition {\it i.e.} $\sin\frac{\theta_1-\theta_2}{2}=0$ maximizes $F_L$.
 Therefore, when $p\geq \dfrac{1}{2}$, $F_L$, like \(F_G\), is also identically equal to $p$ making $S_P=0$. For $0\leq p<\dfrac{1}{2}$,
 $F_L=\dfrac{(1-p)^2}{2-3p}$ and in this region $F_G=1-p$ and therefore, $S_P=\dfrac{(1-p)(1-2p)}{2-3p}$.
 So, finally we have that for the state \(\rho_{ent}\), the shared purity is given by 
\begin{eqnarray}
 S_P &=&  \dfrac{(1-p)(1-2p)}{2-3p},  \quad \quad  0 \leq p < \dfrac{1}{2}, \nonumber \\
     &=&  0,  \quad \quad \quad \quad\quad \quad \quad \quad \phantom{i}  \dfrac{1}{2} \leq p \leq 1. 
     \end{eqnarray}
Clearly, the shared purity vanishes for mixed entangled states. Below we will find however that it may be non-vanishing for separable states. 
We exhibit a plot of the shared purity of the state \(\rho_{ent}\) in Fig. 1.
\begin{figure}[hbpt]
\label{always-entgled}
\vspace{-10pt}
\centerline{
\hspace{-3.3mm}
\rotatebox{-90}{\epsfxsize=6cm\epsfbox{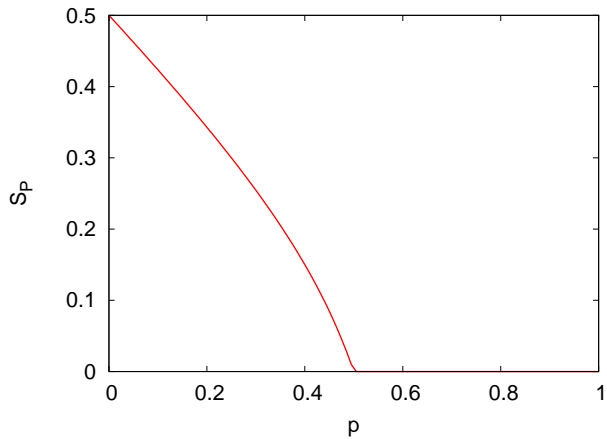}}}
\caption{(Color online) Shared purity (\(S_P\)) with respect to the  parameter $p$, for the  state 
$\rho_{ent}\). The most interesting region, perhaps, is \(\frac{1}{2}\leq p <1\), where the shared purity vanishes, although 
the state has a nonzero entanglement there. Both axes are dimensionless. 
}
\end{figure}
 \subsubsection{Bell mixtures}
Consider next, the mixture of two Bell states, given by 
\begin{equation}
\rho_{Bell}=p|\psi^-\rangle\langle\psi^-|+(1-p)|\psi^+\rangle\langle\psi^+|, \nonumber
\end{equation}
where $|\psi^{+}\rangle=\frac{1}{\sqrt{2}}(|01\rangle+|10\rangle)$ and \(0\leq p \leq 1\). 
Note that the state is entangled for all values of $p$ except $p=1/2$.
 Since the state is already in its spectral decomposition,
$F_G$ is equal to $\max\{p,1-p\}$. $F_L$ is evaluated using the same procedure as given in the preceding example. $F_L$ is
equal to $\dfrac{1}{2}$ for any value of $p$. Therefore,
\begin{eqnarray}
S_P &=& p-\dfrac{1}{2}~~~{\textrm{for}}~~~p\geq\dfrac{1}{2}, \nonumber\\ 
 ~~&=&\dfrac{1}{2}-p~~~{\textrm{for}}~~~p<\dfrac{1}{2}. 
\end{eqnarray}
Just like any quantum correlation measure, shared purity, in this case, is also a mirror reflection with respect to the $p=1/2$ line. This is a result of the local unitary invariance of shared purity (Property 3).
Fig. 2 depicts the behavior of the shared purity for this state. Further Bell mixtures are considered in succeeding examples.
\begin{figure}[hbpt]
\label{fig2}
\vspace{-10pt}
\centerline{
\hspace{-3.3mm}
\rotatebox{-90}{\epsfxsize=6cm\epsfbox{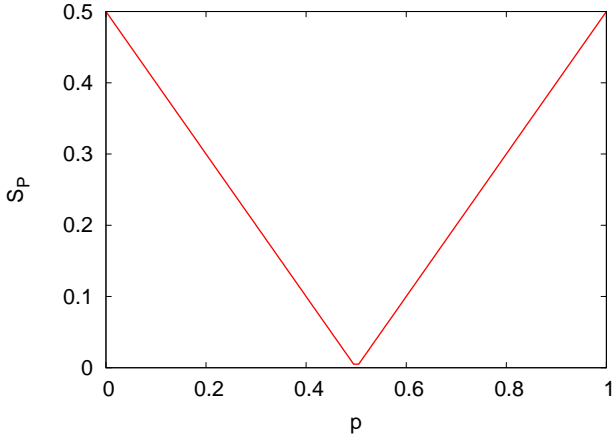}}}
\caption{(Color online) Shared purity (\(S_P\)), as a function of the probability parameter $p$, for the Bell mixture \(\rho_{Bell}\). \(S_P\) is vanishing only at \(p=1/2\).
Both axes are dimensionless.}
\end{figure}
 \subsubsection{Admixtures of pure states with noise}
\begin{figure}[hbpt]
\vspace{-10pt}
\centerline{
\hspace{-3.3mm}
\rotatebox{-90}{\epsfxsize=6cm\epsfbox{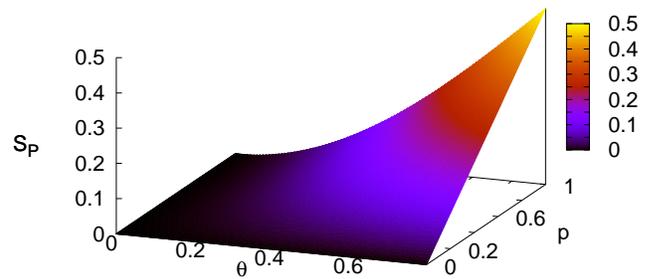}}}
\caption{(Color online) Shared purity (\(S_P\)), as a function of the probability parameter $p$ and the pure state parameter $\theta$ for the shared quantum state
 $\rho_{gen}\). \(\theta\) is measured in radians. Other axes are dimensionless.
}
\label{fig4}
\end{figure}

The mixture of an arbitrary pure state with white noise is given by 
\begin{equation}
\rho_{gen}=p|\psi\rangle\langle\psi|+(1-p)\dfrac{1}{2}I\otimes\dfrac{1}{2}I, \nonumber
\end{equation}
where $|\psi\rangle=\cos\theta|00\rangle+\sin\theta|11\rangle$ with $0\leq\theta\leq\pi/4$, \(0\leq p \leq 1\), and \(I\) denotes the 
identity operator on the qubit Hilbert space. Using Theorem 1, 
we obtain 
$F_G=\dfrac{3p+1}{4}$, while $F_L=pf(\theta)+\dfrac{1-p}{4}$, where $f(\theta)=\max\{\cos^2\theta,\sin^2\theta\}$.
Therefore, the shared purity in this case is given by 
\begin{equation}
 S_P=p(1-f(\theta)). 
\end{equation}
Fig.~\ref{fig4} depicts the behavior of $S_P$ with respect to $\theta$ and the mixing parameter $p$. The state reduces to the Werner state \cite{Rfwerner}
when $\theta=\frac{\pi}{4}$, and the shared purity, then, reduces to $\frac{p}{2}$, which is non-zero for the entire range of $p$ except $p=0$.
Note that the Werner state is entangled for $p>1/3$, while quantum discord and quantum work-deficit are non-vanishing for $p>0$. 
Hence, similar to the information-theoretic quantum correlation measures like quantum discord and quantum work deficit, 
 shared purity can be positive for separable states.
 \subsubsection{Multipartite mixed states}
Let us now consider an $N$-party Greenberger-Horne-Zeilinger  state \cite{GHZ}, mixed with white noise, in $\left(\mathbb{C}^{d}\right)^{\otimes N}$. The state is given by 
\begin{equation}
\rho_{GHZ_N}=p|\psi\rangle\langle\psi|+(1-p)\left(\dfrac{1}{d}I_d\otimes\ldots\otimes\dfrac{1}{d}I_d\right), \nonumber
\end{equation}
where
\begin{equation}
 |\psi\rangle=\dfrac{1}{\sqrt{d}}(|0_1\ldots0_N\rangle+\ldots+|(d-1)_1\ldots (d-1)_N\rangle), \nonumber
\end{equation}
and \(0\leq p\leq 1\). Here, \(I_d\) denotes the identity operator on \(\mathbb{C}^d\), and \(\{|i_j\rangle\}_{i=0}^{d-1}\) for \(j=1,\ldots, N\) forms an orthonormal basis in the Hilbert space of the \(j\)th particle.
In this case, $F_G=p+\dfrac{1-p}{d^N}$, and $F_L$ is given by
\begin{equation}
F_L= \left(\dfrac{p}{d}+\dfrac{1-p}{d^N}\right). 
\end{equation}
Therefore, the shared purity is given by
\begin{equation}
 S_P=p\left(1-\dfrac{1}{d}\right). 
\end{equation}
Note that the shared purity never vanishes 
except at $p=0$.

\section{Monogamy properties of shared purity}
\label{mg}

Certain, and certainly not all, properties of shared quantum systems are known to be monogamic in nature \cite{Akekert, Vcoffman, Rprabhu, score, Salinik}. 
Consider a multiparty quantum system in a state \(\rho_{1\ldots N}\), and a 
certain two-party physical quantity \(\mathcal{Q}\). This quantity will be monogamic 
if a high amount of \(\mathcal{Q}(\rho_{12})\) implies that neither the party 1 nor the party 2 will 
be able to share a substantial amount of \(\mathcal{Q}\) with any other party. Here, 
\(\rho_{12} = \mbox{tr}_{3\ldots N}\rho_{1\ldots N}\). 
However, sharing of classical correlations of a multiparty quantum system does not have any such restriction. 
We begin with the following result which shows that shared purity is indeed a quantum property of shared systems.\\\\
\noindent \textbf{Theorem 4.} If two quantum systems, irrespective of their dimensions, have the maximal amount of shared purity, they cannot share any purity with any third quantum system.\\
\noindent\texttt{Proof:} Consider a three-party system \(\rho_{123}\) in \(\mathbb{C}^{d_1} \otimes \mathbb{C}^{d_2} \otimes \mathbb{C}^{d_3}\), 
such that \(\rho_{12} = \mbox{tr}_3\rho_{123}\) has \(S_P = 1 - 1/d\), where \(d=\min\{d_1,d_2\}\). Note that Corollary 3.1 implies that \(1-1/d\) is the maximal shared purity possible in \(\mathbb{C}^{d_1} \otimes \mathbb{C}^{d_2}\), 
and that it is attained 
by a pure state (any maximally entangled state). 
We now show that there are  no mixed states that also attains that maximal value. 
Let us assume that  \(\rho_{12}\) is nonpure, so that its maximal spectral eigenvalue \(\lambda_r\) is strictly less than unity. Then, by Theorem 3, the shared purity for \(\rho_{12}\) is \(\leq \lambda_r(1-1/d)\), which 
is strictly less than \(1-1/d\). This is a contradiction, implying that \(\rho_{12}\) is pure. 
However, if \(\rho_{12}\) is pure, \(\rho_{123}\) must be of the form \(\rho_{12} \otimes \rho_{3}\). Correspondingly, the state of 1 and 3 will be of the form \(\rho_1 \otimes \rho_3\), so that by Property 5, 
its shared purity must vanish. Similarly, the shared purity between 2 and 3 must also vanish. \hfill \(\blacksquare\)\\\\

Physical quantities of shared systems for which Theorem 4 holds may be termed as ``qualitatively monogamous'', and shared purity is one of them. 
In order to make this statement more quantitative, the relation \cite{Vcoffman, Rprabhu, score, Salinik}
\begin{equation}
\mathcal{Q}(\rho_{12})+ \mathcal{Q}(\rho_{13}) \leq \mathcal{Q}^{1:23}(\rho_{123}) 
\label{monoeq1}
\end{equation}
is usually considered, where \(\mathcal{Q}^{1:23}(\rho_{123})\) denotes the value of \(\mathcal{Q}\) of the state \(\rho_{123}\) in the 1:23 partition. A quantum state \(\rho_{123}\) is said to be monogamous 
with respect to the bipartite physical quantity \(\mathcal{Q}\), if the relation in (\ref{monoeq1}) is satisfied. 
Let us therefore consider the relation 
\begin{equation}
 S_P(\rho_{12})+S_P(\rho_{13})\leq S_P^{1:23}(\rho_{123}), 
\label{monoeq}
\end{equation}
and find the states for which this relation is satisfied and whether there are violations of it.

Let us first consider the generalized GHZ state, given by 
\begin{equation}
 |\psi\rangle^G_{GHZ}=\cos\theta|000\rangle+e^{i\phi}\sin\theta|111\rangle, \nonumber
\end{equation}
where $\theta\in[0,\pi]$ and $\phi\in[0,2\pi)$.
The two-party reduced density matrices are
\begin{equation}
\rho_{1j}^G=\cos^2\theta|00\rangle\langle00|_{1j}+\sin^2\theta|11\rangle\langle11|_{1j}, \nonumber
\end{equation}
where $j\in\{2,3\}$. Since the state is classically correlated, $S_P(\rho_{1j}^G)=0~\forall j$, as shown in Sec.~\ref{ms}. 
Now since $|\psi\rangle_{GHZ}^G$ is a pure state, $F_G^{1:23}=1$, 
while $F_L^{1:23}=\max \{\cos^2\theta,\sin^2\theta\}$. Therefore, 
\begin{equation}
S_P^{1:23}(|\psi\rangle^G_{GHZ})=1-\max \{\cos^2\theta,\sin^2\theta\}. 
 \end{equation}
Hence, the monogamy condition, \textit{viz.} Eq.~(\ref{monoeq}), is always satisfied for the generalized GHZ states.

\begin{figure}[hbpt]
\vspace{-10pt}
\centerline{
\hspace{-3.3mm}
\rotatebox{-90}{\epsfxsize=6cm\epsfbox{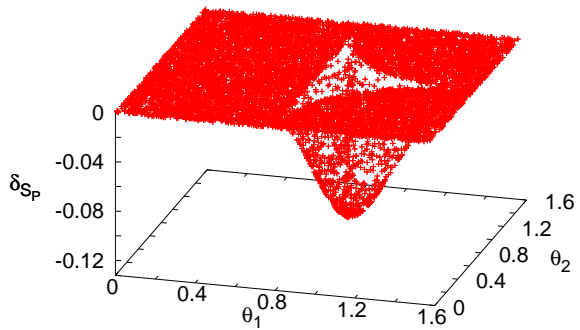}}}
\caption{(Color online) Monogamy score for shared purity of the generalized W states. 
$\delta_{S_P}$, plotted on the vertical axis, is positive for all $\theta_1$ and $\theta_2$,  implying that the generalized W states are always non-monogamous. The monogamy score is a 
dimensionless quantity, while \(\theta_1\) and \(\theta_2\) are measured in radians. For every choice of \((\theta_1, \theta_2)\), the values of \(\phi_1\) and
\(\phi_2\) are chosen randomly.}
\label{gnrl-W}
\end{figure}

Next, let us  consider the generalized W state, given by
\begin{eqnarray}
 |\psi\rangle^G_{W}&=&\sin\theta_1\cos\theta_2|001\rangle+\sin\theta_1 \sin\theta_2 e^{i\phi_1}|010\rangle+ \nonumber \\
&& +\cos\theta_1 e^{i\phi_2}|100\rangle. \nonumber
\end{eqnarray}
In this case, $F_L^{1:23}=\max \{\sin^2\theta,\cos^2\theta\}$, and therefore 
$S_P^{1:23}=\min \{\sin^2\theta,\cos^2\theta\}$. 
On the other hand, 
\begin{eqnarray}
F_G(\rho_{12}^W)&=&\max \{\sin^2\theta\cos^2\phi,1-\sin^2\theta\cos^2\phi\}\nonumber \\ 
F_G(\rho_{13}^W)&=&\max \{\sin^2\theta\sin^2\phi,1-\sin^2\theta\sin^2\phi\}.
\end{eqnarray}
The local fidelities of the local density matrices, $\rho_{1j}^W$, of the state \(|\psi\rangle^G_{W}\), 
where $j\in\{2,3\}$, are maximized numerically, by generating $2\times10^{4}$ random
states.  We find that the generalized W states are always polygamous with respect to shared purity. In Fig.~\ref{gnrl-W}, 
we plot the ``shared purity monogamy score'', defined as~\cite{score}
\begin{equation}
 \delta_{S_P}=S_P(\rho_{1:23})-(S_P(\rho_{12})+S_P(\rho_{13})) 
\end{equation}
as a function of $\theta_1$ and $\theta_2$.
Considering only generalized GHZ and generalized W states, the monogamy properties of shared purity are very similar to those of quantum discord~\cite{Rprabhu, score}. 
The same is, however, not true for bigger classes of states, as we 
will now find.

Let us now consider the GHZ-class~\cite{Wdur} and the W-class states~\cite{Wdur}. These disjoint sets, when taken together, encompass the complete set of three-qubit pure states.
The normalized GHZ-class states can be represented by
\begin{equation}
 |\psi\rangle_{ABC}=\sqrt{K}\left(\cos\delta|000\rangle+e^{i\phi}\sin\delta|\xi_A\rangle|\xi_B\rangle|\xi_C\rangle\right), \nonumber
\end{equation}
where
\begin{eqnarray}
 |\xi_A\rangle&=&\cos\theta_1|0\rangle+\sin\theta_1|1\rangle, \nonumber \\ 
 |\xi_B\rangle&=&\cos\theta_2|0\rangle+\sin\theta_2|1\rangle, \nonumber \\
 |\xi_C\rangle&=&\cos\theta_3|0\rangle+\sin\theta_3|1\rangle, \nonumber 
\end{eqnarray}
and $K=\left(1+2\cos\delta\sin\delta\cos\theta_1\cos\theta_2\cos\theta_3\cos\phi\right)^{-1} \in (1/2,\infty)$ is a normalization factor.
The ranges for the five parameters are $\delta\in(0,\pi/4]$, $\theta_1,\theta_2,\theta_3\in(0,\pi/2]$ and $\phi\in[0,2\pi)$.
The monogamy relation, in Eq.~(\ref{monoeq}), is checked numerically by randomly choosing $10^{4}$ states from the GHZ-class states. We  
find that 54.36\% states are non-monogamous. 
Next, we consider the normalized W-class states, given by
\begin{equation}
 |\psi\rangle_{ABC}=a|001\rangle+b|010\rangle+c|100\rangle+d|000\rangle \nonumber
\end{equation}
where
\begin{eqnarray}
a&=&\sin\phi_1 \sin\phi_2 \sin\phi_3, \nonumber \\
b&=&\cos\phi_1 \sin\phi_2 \sin\phi_3, \nonumber \\
c&=&\cos\phi_2 \sin\phi_3, \nonumber \\
d&=&\cos\phi_3, \nonumber
\end{eqnarray}
with $\phi_1, \phi_2 \in \left(0,\pi/2\right)$ and $\phi_3 \in \left(0,\pi/2\right]$.
We find that 82.61\% of the $10^{4}$ randomly chosen states from the W-class turn out to be non-monogamous. 
In case of quantum discord, when the measurement is performed on the second party, all W-class states are non-monogamous \cite{Rprabhu}. 
We now consider the monogamy relation for the square of shared purity (cf. \cite{Salinik}), {\it i.e.}, we introduce the score
\begin{equation}
 \delta^{(2)}_{S_P}=S_P^2(\rho_{1:23})-(S_P^2(\rho_{12})+S_P^2(\rho_{13})). 
\end{equation}
In this case, 37.66\% of the randomly chosen states from the GHZ-class remain non-monogamous, while among the W-class states, 
62.39\% continue to be non-monogamous.

\section{Application: Detecting criticality in quantum spin models by shared purity}
\label{qpt}
In this section, we show that shared purity can be applied to detect cooperative phenomena in quantum many-body systems. We will find that 
the scaling exponents of shared purity near critical points are different 
from those of quantum correlation measures. 
More specifically, we 
study the scaling behavior of the shared purity in the one-dimensional anisotropic quantum XY models \cite{Barouch-McCoy}. 
Such models can be simulated using  ultracold gases  in a controlled way in the laboratories (see \cite{Strotzky, Mlewenstein}), and is also known to describe 
Hamiltonians of materials~\cite{Mschechter}.
We therefore consider a  system of  $N$ quantum spin-1/2 particles, arranged on a ring, and described by the anisotropic quantum XY model. 
The Hamiltonian corresponding to which is given by
\begin{equation}
\label{eq_XY_H}
 H_{XY} = \frac{J}{2} \left(\sum_{i=1}^{N} (1 + \gamma) \sigma^x_i \sigma^x_{i+1} + (1 - \gamma) \sigma^y_i \sigma^y_{i+1}\right) + h \sum_{i=1}^{N} \sigma_i^z,
\end{equation}
where $J$ is the coupling constant for the nearest neighbor interaction, $\gamma \in (0,1]$ is the anisotropy parameter, \(\sigma\)'s are the Pauli spin matrices, 
and \(h\) represents the external transverse magnetic field applied across the system. Periodic boundary condition is assumed here, so that 
\(\vec{\sigma}_{N+1} \equiv \vec{\sigma}_{1}\).
$\gamma=1$ corresponds to the transverse Ising model.
The above Hamiltonian can be diagonalized by applying Jordan-Wigner, Fourier, and Bogoliubov transformations~\cite{Barouch-McCoy}
successively. At zero temperature, it undergoes 
a quantum phase transition driven by the transverse magnetic field. Such transitions have been detected by using 
different order parameters~\cite{Barouch-McCoy,Mefisher}, including quantum correlation measures like 
concurrence~\cite{Aosterloh}, 
geometric measures~\cite{Ashimony,Asde,Abiswas}, and quantum discord~\cite{Rdillenschneider}.

\begin{figure}[hbpt]
\vspace{-10pt}
\centerline{
\hspace{-3.3mm}
\rotatebox{-90}{\epsfxsize=6cm\epsfbox{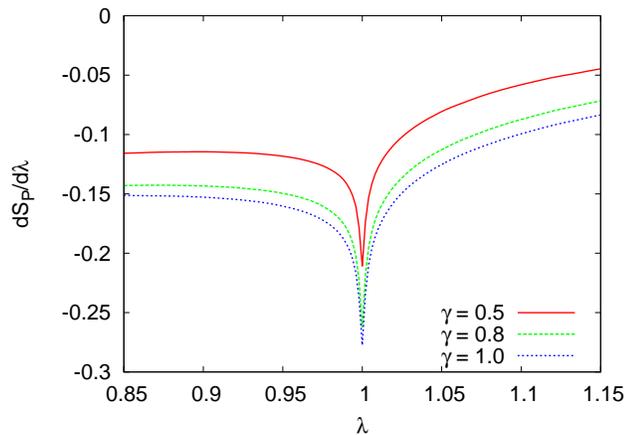}}}
\caption{(Color online) Shared purity detects quantum phase transition. The derivative of $S_P$, \emph{i.e.} $\frac{dS_P}{d\lambda}$, is plotted 
against $\lambda$ for the one-dimensional anisotropic quantum XY model in the thermodynamic limit.
We have chosen three different anisotropy values, {\it viz.} $\gamma=0.5$, 0.8, 1.0. Both axes represent dimensionless parameters.}
\label{fig-aniso}
\end{figure}

We now investigate the behavior of the shared purity of the nearest neighbor density matrix of the ground state near the known quantum critical point at 
$\lambda=\dfrac{h}{J}=1$~\cite{Barouch-McCoy}.
The nearest neighbor two-body density matrix corresponding to the ground state of the XY Hamiltonian, Eq.~(\ref{eq_XY_H}), represented by $\rho_{AB}$, 
can be written~\cite{Barouch-McCoy} in terms of the classical two-site correlations and the average magnetization
in the direction of the external magnetic field {\it i.e.} the $z$-direction. The density matrix, $\rho_{AB}$, in the thermodynamic limit of \(N\to \infty\), is given by\\\\
\hspace*{1cm}$\small{\rho_{AB}=\begin{pmatrix} 
  \alpha_{+}+\dfrac{M_z}{2} & 0 & 0 & \beta_{-}\\ 
  0 & \alpha_{-} & \beta_{+} & 0\\
  0 & \beta_{+} &  \alpha_{-} & 0\\
  \beta_{-} & 0 & 0 & \alpha_{+}-\dfrac{M_z}{2}
\end{pmatrix}}$\\\\
where $\alpha_{\pm}=\dfrac{1}{4}(1\pm T_{zz})$, $\beta_{\pm}=\dfrac{T_{xx}\pm T_{yy}}{4}$ with $T_{ij}=\mbox{tr}(\sigma_i\otimes\sigma_j\rho_{AB})$
and $M_z=\mbox{tr}(\sigma_z \rho_A)$.
The correlations and transverse magnetization, for the zero-temperature state, are given by~\cite{Barouch-McCoy} 
\begin{eqnarray}
T^{xx}(\tilde{h})&=&G(-1, \tilde{h}),\nonumber\\
T^{yy}(\tilde{h})&=&G(1, \tilde{h}),\\
T^{zz}(\tilde{h})&=& [M^z(\tilde{h})]^2-G(1, \tilde{h})G(-1, \tilde{h}),\nonumber
\label{eq:Tijeq}
\end{eqnarray}
where 
$G(R, \tilde{h})$ (for $R =\pm 1$) are 
\begin{eqnarray}
G(R, \tilde{h}) &=&  \frac{1}{\pi}\int^\pi_0d\phi \frac{1} {\Lambda(\tilde{h})} \nonumber\\
&\times &(\gamma \sin(\phi R)\sin \phi - \cos\phi (\cos\phi -\tilde{h})) \nonumber\\
\end{eqnarray}
and 
\begin{eqnarray}
M^z(\tilde{h})) &=& -\frac{1}{\pi} \int_0^\pi d\phi \frac{ (\cos\phi -\tilde{h})}{\Lambda(\tilde{h})}. \nonumber \\
 \label{eq:Mzeq}
\end{eqnarray}
Here 
\begin{equation}
\Lambda(x)= \left\{\gamma^2\sin^2\phi~+~[x-\cos\phi]^2\right\}^{\frac{1}{2}},
\end{equation}
and 
\begin{equation}
 \tilde{h} = \frac{h}{J}. 
\end{equation}
Note that \(\tilde{h}\)
is a 
dimensionless variable.
$F_G$ corresponds to the maximum eigenvalue of the density matrix $\rho_{AB}$. 
$F_L$ is obtained by numerical maximization of the density matrix $\rho_{AB}$ with respect to the product states in $\mathbb{C}^2 \otimes \mathbb{C}^2$.
 We plot the derivative of $S_P$ with respect to $\lambda$, \emph{i.e.} $\frac{dS_P}{d\lambda}$, against $\lambda$, in Fig.~\ref{fig-aniso},
for different values of the anisotropy parameter $\gamma$.
The divergence of the derivative at $\lambda=\lambda_c \equiv 1$ clearly signals the quantum phase transition in this model, indicating that 
shared purity can be used as a physical quantity to detect 
quantum phase transitions. \\

\begin{figure}[hbpt]
\vspace{-10pt}
\centerline{
\hspace{-3.3mm}
\rotatebox{0}{\epsfxsize=8cm\epsfbox{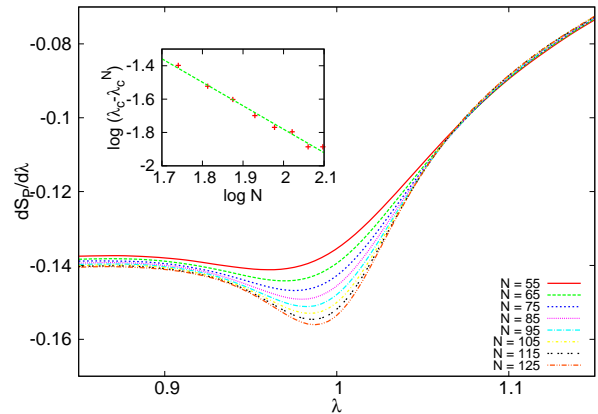}}}
\caption{(Color online) Finite-size scaling analysis for shared purity. The derivative of $S_P$ with respect to $\lambda$ is plotted against 
$\lambda$ for the one-dimensional anisotropic quantum XY model
for finite numbers of particles, for $\gamma=0.8$. The behavior is similar for other anisotropy values. The inset shows the scaling analysis 
for finite \(N\), which is a log-log plot between \(\log_{10}N\) (the horizontal axis) and 
\(\log_{10}(\lambda_c - \lambda_c^N)\) (on the vertical axis), where \(\lambda_c^N\) is the value of \(\lambda\) at which \(\frac{dS_P}{d\lambda}\) 
attains a minimum, for a system of \(N\) spins. All axes are dimensionless.
The base 10 of the logarithm is not displayed in the inset for simplicity.}
\label{fig-aniso-finite}
\end{figure}

\noindent \textbf{Finite-size scaling.} We now perform a finite-size scaling analysis for the shared purity near the critical point. Such scaling 
analysis will lead us to obtain the scaling exponents of shared purity in this 
model. Another importance of such  analysis is that it helps us to understand the viability of detecting the critical point in finite-sized 
systems, which can be built by using ultracold gas systems. 
We have calculated the shared purity of nearest neighbor spins for finite chains consisting of \(N\)  spins, with $N=55,65,75,85,95,105,115,125$. 
By performing the scaling analysis, we find that the 
point of divergence approaches $\lambda=\lambda_c$ as $N^{-1.4}$ {\it i.e.} 
\begin{equation}
 \lambda=\lambda_c+kN^{-1.4}
\end{equation}
where $k$ is a dimensionless constant. We plot the derivative of
shared purity with respect to $\lambda$ for a finite number of spins along with the scaling in Fig.~\ref{fig-aniso-finite}.

It is to be noted here that the finite-size scaling exponent obtained for shared purity is different from those for other quantum correlation measures. 
In particular, for the entanglement measure called
concurrence, finite-size analysis led to an approach to the point of divergence as \(N^{-1.87}\) \cite{Aosterloh}, while for the information-theoretic 
quantum correlation measure called 
quantum discord, the same gives the point of divergence as \(N^{-1.28}\) \cite{exponent-discord}.

\vspace*{1cm}
\vspace*{1cm}

\section{Discussion}
\label{concl}
Entanglement, and in recent years, information-theoretic quantum correlations, have been one of the most important pillars in the grand edifice of 
quantum information  of 
shared quantum systems. 
There have, however, been indications that this may not be the entire picture, and that there may be important resources that are employed by nonclassical 
phenomena in the domain of 
shared quantum systems. We have conceptualized and provided a measure of shared quantum systems, which we have called shared purity, and which we have shown 
to be independent of quantum correlations, qualitatively and 
quantitatively. The concept is based on the maximal fidelity of the shared quantum state to certain sets of pure quantum states. 
The measure is defined for an arbitrary 
state (pure or mixed) of an arbitrary number of parties in arbitrary dimensions. For pure shared states, the shared purity reduces to the geometric  
measures of entanglement. However, for mixed states, the measure is shown to exhibit a drastically different behavior from any measures of quantum correlation. 
The measure is a difference between two quantities, which we have called the global and local fidelities. The global fidelity is shown to be expressible in 
closed analytic form. The local fidelity is also 
calculated analytically for several paradigmatic classes of mixed shared quantum states. Efficient numerical procedures are possible in other cases. 

We have shown that the measure is qualitatively monogamous, in that a maximal amount of shared purity between two quantum systems rules out 
the existence of any shared purity with these 
systems of any third party. This is very unlike classical correlations, and ensures the quantum nature of the quantity. 
Like many other measures of entanglement and information-theoretic quantum correlations, shared purity can be 
both monogamous and non-monogamous, when quantitatively probed. We have performed this quantitative analysis for all three-qubit pure states. In particular, 
we have found that 
all generalized W states violate the monogamy relation for shared purity while all generalized Greenberger-Horne-Zeilinger states  satisfy the same. 

Finally, we have applied the shared 
purity to analyze the quantum anisotropic XY spin model and have found that it can faithfully signal the quantum phase transition present in this model. Moreover, we have 
performed a finite-size scaling analysis, and have obtained the scaling exponent of shared purity in this model. Interestingly, the 
scaling exponent is different from those of 
concurrence, an entanglement measure, and quantum discord, an information-theoretic quantum correlation.

\end{document}